\documentclass[lettersize,journal]{IEEEtran}

\usepackage{cite}

\usepackage{graphicx}

\usepackage{amsmath}
\usepackage{wasysym}
\usepackage{amssymb}

\usepackage{subcaption}
\usepackage{multicol}
\usepackage{tabularx}

\usepackage{algorithmic}
\usepackage{comment}

\usepackage{diagbox}

\usepackage{array}

\usepackage{fixltx2e}

\usepackage{dblfloatfix}

\usepackage{url}

\usepackage[dvipsnames]{xcolor}

\begin{document}

\title{Collaborative Semantic Communication for Edge Inference}


\author{\IEEEauthorblockN{Wing Fei Lo,
Nitish Mital,~\IEEEmembership{Member,~IEEE,}
Haotian Wu,~\IEEEmembership{Graduate Student Member,~IEEE,}\\
Deniz G\"{u}nd\"{u}z,~\IEEEmembership{Fellow,~IEEE}
}
\thanks{The authors are with the Department of Electrical and Electronic Engineering, Imperial College London, London SW7 2AZ, U.K. (e-mail: haotian.wu17@imperial.ac.uk).}

}

%



\IEEEtitleabstractindextext{%
\begin{abstract}
We study the collaborative image retrieval problem at the wireless edge, where multiple edge devices capture images of the same object from different angles and locations, which are then used jointly to retrieve similar images at the edge server over a shared multiple access channel (MAC). We propose two novel deep learning-based joint source and channel coding (JSCC) schemes for the task over both additive white Gaussian noise (AWGN) and Rayleigh slow fading channels, with the aim of maximizing the retrieval accuracy under a total bandwidth constraint. The proposed schemes are evaluated on a wide range of channel signal-to-noise ratios (SNRs), and shown to outperform the single-device JSCC and the separation-based multiple-access benchmarks. We also propose a channel state information-aware JSCC scheme with attention modules to enable our method to adapt to varying channel conditions.
\end{abstract}

}

\maketitle

\IEEEdisplaynontitleabstractindextext

%

\section{Introduction}
%
%
%
%
\IEEEPARstart{I}{n} recent years, machine learning tasks at the wireless edge have been studied extensively in the literature, including distributed and remote inference problems over wireless channels \cite{gunduz2020communicate,shao2021learning,9614039}. In distributed and remote inference problems, it is often assumed that centrally trained models, e.g. deep neural networks (DNNs) are employed across multiple distributed nodes, which have limited communication resources. 
Particularly in image retrieval, images of an object or a person taken by edge devices are used to identify the images of the same object or person, taken by different cameras, from different angles, and at different times, in a gallery database. 
Note that for image retrieval, unlike most conventional classification or inference problems, which can be carried out locally at the edge device, remote inference is essential even if the edge devices have unlimited computational power, as the gallery database is only available at the edge server. On the other hand, due to latency and bandwidth constraints, sending the whole image over a noisy wireless channel is not feasible. Instead, learning-based feature extraction is done at the edge, and only the most relevant features of the source image, representing the semantic content of the image, are sent to the edge server over the wireless channel. This calls for semantic communication, since the inference classes are not pre-defined, and the edge server must infer the similarity of the semantics of the image captured by the edge device with those of the images in the gallery database. This approach is also called \textit{goal-oriented communication} in the \textit{semantic communication} literature  \cite{gunduz2022beyond}.

In \cite{jscc_ae}, both separation-based and joint source channel coding (JSCC) approaches have been studied for feature transmission in remote image retrieval. While Shannon's separation theorem \cite{shannon} states that separating source and channel coding can achieve asymptotic optimality, this theorem breaks down in finite block-lengths. We typically have much more stringent latency constraints on edge inference applications compared to the delivery of images or videos; hence, our interest is in very short blocklengths, where separation typically has very poor performance. An autoencoder-based JSCC (JSCC-AE) scheme is proposed in \cite{jscc_ae}, and it is shown to outperform its digital counterpart under all channel conditions. 

In this paper, we study the collaborative re-identification (ReID) problem, where two edge devices capture images of the same scene {and communicate with the edge server, in a distributed manner, to predict the image identity based on similar images in a gallery database. The distributed nature of the problem poses unique challenges, where the edge devices must ``collaborate'' implicitly to derive the relevant semantic information from their respective images of the scene, in a manner which complements the other and therefore improves the communication or inference accuracy at the receiver. We highlight that such collaboration is implicit, and not explicit where the edge devices would share messages with each other.}

\begin{figure}[!t]
\centering
\includegraphics[width=1\linewidth]{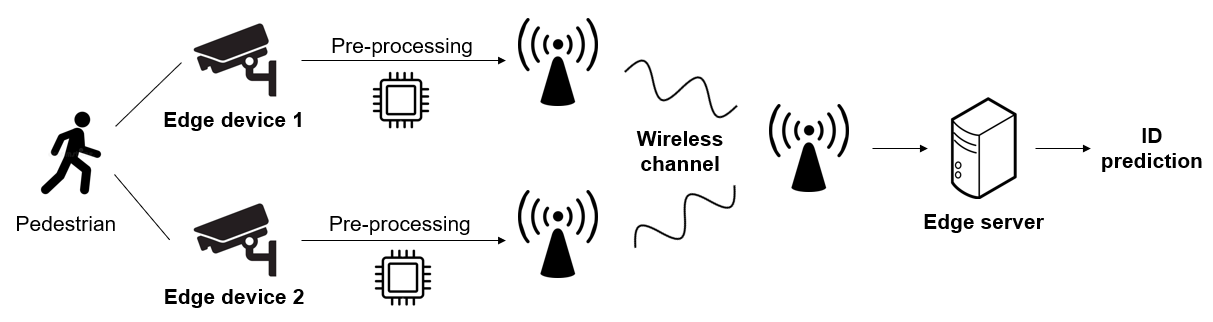}
\caption{Illustration of the two-device collaborative image retrieval problem at the wireless edge.}
\label{fig:person_reid}
\end{figure}


The goal of this paper is to develop a deep learning-based JSCC scheme for the two-device scenario, which maximizes the accuracy of the retrieval task while communicating over a shared multiple access channel (MAC). {To explore different transmission schemes for the multi-source collaborative edge inference,} we first consider an orthogonal multiple access (OMA) scheme employing time division multiple access (TDMA) with distributed JSCC, and show that it outperforms the schemes in \cite{jscc_ae}, as well as a conventional separate source-channel coding scheme, where each device transmits a quantized version of its features to the receiver using capacity-achieving channel codes. In addition, we study an alternative non-orthogonal multiple access (NOMA) approach. Benefits of NOMA transmission in various distributed inference and training problems have recently received significant interest \cite{amiri2020machine, Yilmaz:ISIT:22, Krouka:Globecom:22}. In {the NOMA} approach, our goal is to exploit the superposition property of the wireless medium, and the features transmitted as analog values over the shared wireless channel get aggregated ``over-the-air'', {thus boosting the signal associated with the common semantic information in the two transmitted signals, correlated with the common identity viewed by the edge devices}. We evaluate these schemes on the additive white Gaussian noise (AWGN) and Rayleigh slow fading channels. Inspired by the attention mechanism in adaptive JSCC \cite{jscc_attention,wu2022vision,wu2022channel}, we also propose an SNR-aware scheme for the AWGN channel to adjust the networks depending on the SNRs. Our main contributions can be summarized as follows:

\begin{itemize}
\item To the best of our knowledge, this is the first paper to study collaborative inference among edge devices for joint retrieval. We propose two new collaborative JSCC schemes for OMA and NOMA transmissions, and show the superiority of the latter.

\item
We construct and analyze DNN architectures for a channel state information (CSI)-aware JSCC scheme (SNR-aware and channel fading-aware),  where a single network is trained to exploit the channel state information for channel equalization and SNR-adaptation.

\end{itemize}

\section{Related work}
\subsection{Image retrieval} 
Image retrieval task aims to improve the quality of identity recognition. Given a query image, an image retrieval model assesses its similarities with gallery images, and matches it to the `nearest' ones. Performance can be evaluated through top-1 retrieval accuracy \cite{cmc}.  Image retrieval task has received significant attention in recent years thanks to the tremendous success of deep learning technologies \cite{personreid}. 

\subsection{Remote inference at the wireless edge} 
{With the rapid growth of machine intelligence and the associated machine-to-machine communications, the goal of emergent communication systems is shifting towards making accurate inferences about a remote signal rather than reconstructing it \cite{gunduz2020communicate}, unlike conventional communication systems which are designed to serve data packets without regarding the content of the packets or the task at the receiver. Therefore, remote inference problems are attracting significant interest in the context of the emerging semantic communication paradigm \cite{gunduz2022beyond}. Literature on joint edge-device inference mostly focus on a rate-limited scenario \cite{digital_2,digital_1}, while ignoring channel effects. Jankowski et al.\cite{jscc_ae} proposed a JSCC transmission scheme for image retrieval, showing a marked improvement over previous works based on digital schemes.}  

\subsection{Multi-device collaborative learning}
{Existing multi-device collaborative algorithms mainly focus on signal transmission\cite{bian2022deep}, classification tasks \cite{shao2022task}, visual question answering\cite{xie2022task}, and multi-agent coordination \cite{lotfi2022semantic}}. 
Shao et al. \cite{shao2022task} propose a deterministic distributed information bottleneck (DDIB) principle for distributed feature encoding. Different from previous work, our paper studies collaborative inference over the wireless edge, in which the effects of a wireless channel are considered.

\section{System model}
We consider two transmitters, each having access to images of the same object taken by a different camera. We denote the image observed by transmitter $i$ by $\mathbf{s}_i \in \mathbb{R}^{p}$, $i = 1, 2$. Transmitter $i$ employs an encoding function $\mathcal{E}_i: \mathbb{R}^{p} \rightarrow \mathbb{C}^{q}$, where $\mathbf{x}_i = \mathcal{E}_i (\mathbf{s}_i) \in \mathbb{C}^q$ {and $\mathbf{x}_i$ is subject to the power constraint as: $\frac{1}{q}\vert\vert \mathbf{x}_i \vert\vert^2_2 \leq 1$} . Here, $q$ represents the available channel bandwidth. 
The decoder function $\mathcal{D}: \mathbb{C}^{q} \rightarrow \mathbb{D}$ is employed at the receiver, where $\mathbb{D}\equiv \{1, 2, \ldots, D \}$, and $D$ is the size of the database, maps the received signal $\mathbf{y}$ to the result of the retrieval task.

\textbf{Channel model:}
Devices transmit their signals over a MAC. 
{The received signal is given by $\mathbf{y} = h_1\mathbf{x}_1 + h_2\mathbf{x}_2 + \mathbf{z}$, where $\mathbf{z} \in \mathbb{C}^q$ is the additive noise vector, assumed to be independent and identically distributed (i.i.d.) according to the complex normal distribution $\mathcal{CN}(0,\sigma_z^2)$. For the AWGN channel, we set $h_1=h_2=1$.} We also consider a slow fading MAC, where the fading coefficients $h_1$ and $h_2\in \mathbb{C}$, assumed to remain constant during each retrieval task, but changes across tasks in an i.i.d. fashion sampled from $\mathcal{CN}(0,\sigma_h^2)$.


We will consider and compare three alternative transmission schemes, separation-based transmission, JSCC with OMA, and JSCC with NOMA, as well as the single-user benchmark\cite{jscc_ae}. 

\subsection{Separate Digital Transmission}
In the digital scheme, transmitter $\mathcal{E}_i$ extracts a {semantic feature vector} $\mathbf{v}_i \in \mathbb{R}^r$ from the source $\mathbf{s}_i$, which is quantized to $\tilde{\mathbf{v}}_i \in \mathbb{Z}^r$, and then mapped to a channel codeword $\mathbf{x}_i \in \mathbb{C}^q$. The two transmitters transmit their codewords over the MAC.

The receiver first decodes the two channel codewords to recover the quantized {semantic features} $\tilde{\mathbf{v}}_1$ and $\tilde{\mathbf{v}}_2$.
In the asymptotic limit of infinite blocklength, the transmitted codewords can be decoded with a vanishing error probability if the transmission rates are within the capacity of the corresponding channels. In that case, the only source of error in the computation of the desired function is quantization. 
The receiver then performs the retrieval task on the recovered source signals.

\subsection{JSCC}
 {In this scheme, source signals $\mathbf{s}_i\in \mathbb{R}^p, i=1,2,$ are first mapped to semantic feature vectors $\mathbf{v}_i \in \mathbb{R}^r, i=1,2$, which are then mapped to the channel codewords $\mathbf{x}_i \in \mathbb{C}^q$. 
We consider two JSCC schemes:}

 {
\textbf{JSCC with OMA:}
Each transmitter is allocated half the available channel bandwidth, i.e., $\frac{q}{2}$ channel uses. }

 {
\textbf{JSCC with NOMA:}
In this scheme, each transmitter occupies the full channel bandwidth of $q$.} 

 {
In both cases, the receiver first decodes the received signal, using two JSCC decoders $\mathcal{D}_i: \mathbb{C}^q \rightarrow \mathbb{R}^r, i=1,2$, to recover estimates $\hat{\mathbf{v}}_1$ and $\hat{\mathbf{v}}_2$ of the semantic features, and then performs the retrieval task using the recovered semantic features}.

\section{Distributed image retrieval}
In this section, we focus on the image retrieval task,  {which is evaluated by the top-1 retrieval accuracy \cite{cmc}.}

\subsection{Separate Digital Transmission}
Each transmitter consists of a  {semantic feature encoder}, modeled as a ResNet50 \cite{resnet} network, followed by a feature compressor, employing quantization and arithmetic coding  {modules, which are the same as the state-of-the-art pipeline in \cite{jscc_ae}}. The compressed bits are then channel coded and transmitted on the wireless channel. The receiver decodes the received signal to obtain estimates of the quantized  {semantic features}, which are then passed to the image retrieval module.

\textbf{Training strategy:}
We perform end-to-end training for the digital scheme, with the following loss function: $l = \frac{1}{3} (l_{ce_{aux1}} + l_{ce_{main}} + l_{ce_{aux2}}) + \lambda \cdot (\log_2 p(\tilde{\mathbf{v}}_1) + \log_2p(\tilde{\mathbf{v}}_2))$,
where $l_{ce_{aux1}}, l_{ce_{aux2}}, l_{ce_{main}}$ are the cross-entropy losses between the  {identity prediction result} from  {each} classifier (two auxiliary and a main classifier, see Fig. 2) and the ground truth,  {same as {\cite{personreid, jscc_ae}}}. $\log_2 p(\tilde{\mathbf{v}}_1)$ and $\log_2 p(\tilde{\mathbf{v}}_2)$ are entropies of the quantized semantic features,  {same as in \cite{jscc_ae}}.

\subsection{JSCC}\label{sec:jscc}
\begin{figure}[t]
\begin{center}
   \includegraphics[width=1\linewidth]{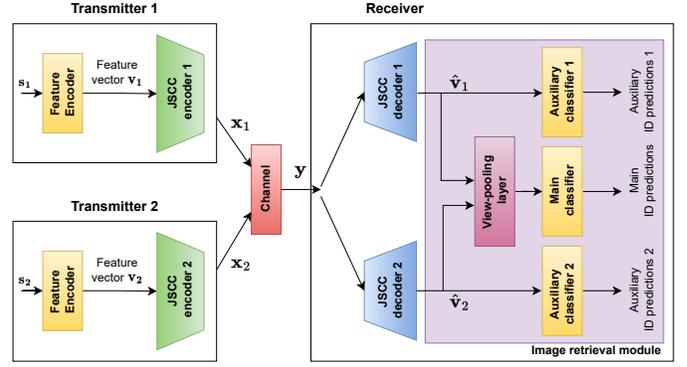}
   \end{center}
   \caption{ {DNN architecture for the JSCC transmission schemes.} }
\label{fig:jscc}
\end{figure}

In this scheme (illustrated in Fig. \ref{fig:jscc}), the feature compressor, quantizer, arithmetic coder, and channel coder at the transmitter, and the channel decoder and arithmetic decoder at the receiver, are replaced by a single autoencoder architecture. The received signal is fed to two  {joint semantic-JSCC decoders}, which decode estimates of the semantic features sent by the two transmitters. Once the semantic features are recovered, they are used for the image retrieval task.

\textbf{Training strategy:}
A three-step training strategy is adopted, which consists of pre-training of the semantic feature encoders ($\mathrm{T_1}$), pre-training of the JSCC autoencoders ($\mathrm{T_2}$), and end-to-end training ($\mathrm{T_3}$). In $\mathrm{T_1}$, the semantic feature encoder is pre-trained, using the average cross-entropy loss function: $l_{cls} = \frac{1}{3} (l_{ce_{aux1}} + l_{ce_{main}} + l_{ce_{aux2}})$. In $\mathrm{T_2}$, the pre-trained semantic feature encoders are frozen, and only the JSCC autoencoders are trained, using the average mean squared error (MSE) loss between the transmitted and reconstructed semantic features:  {${l}_{jscc} = \frac{1}{2} (l_{MSE_{1}} + l_{MSE_{2}}),$}  {where $l_{MSE_{i}}, i=1,2$ is the mean squared error between the transmitted features $\mathbf{v_i}$ and reconstructed semantic features $\mathbf{\hat{v}_i}$ of the $i$-th transmitter.}
In $\mathrm{T_3}$, the whole network is trained jointly, with the loss function in $\mathrm{T_1}$. 

We also propose a CSI-aware architecture variation for AWGN and slow fading channel with CSI at the receiver only (CSIR), where the available CSI (SNR or channel gain) is fed to the model via attention feature (AF) modules \cite{wu2022channel, jscc_attention} inserted before, after and between each layer of the autoencoder. For the AWGN channel, the AF modules at the encoder and decoder scale the intermediate feature maps to adapt to the channel SNR. 
For slow fading with CSIR, the AF modules scale the received signal and the intermediate feature maps by a channel-dependent constant, intuitively playing the role of channel equalization.

\begin{figure*}[!t]
     \centering
     \begin{subfigure}[b]{0.33\textwidth}
         \centering
         \includegraphics[width=0.9\textwidth]{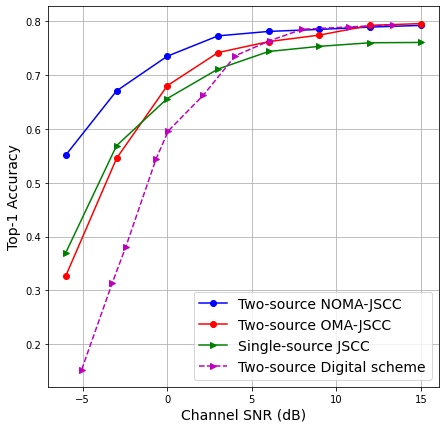}
         \caption{AWGN channel}
         \label{fig:top1_awgn}
     \end{subfigure}%
     \begin{subfigure}[b]{0.33\textwidth}
         \centering
         \includegraphics[width=0.9\textwidth]{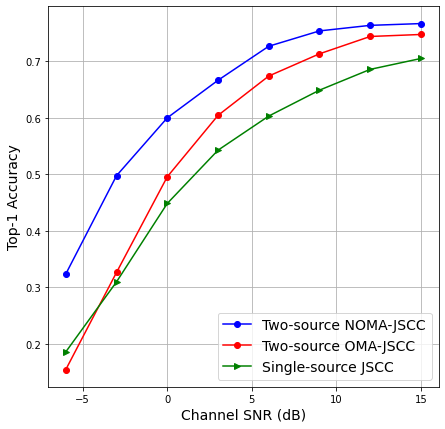}
         \caption{Slow fading channel without CSI}
         \label{fig:top1_nocsi}
     \end{subfigure}%
     \begin{subfigure}[b]{0.33\textwidth}
         \centering
         \includegraphics[width=0.9885\textwidth]{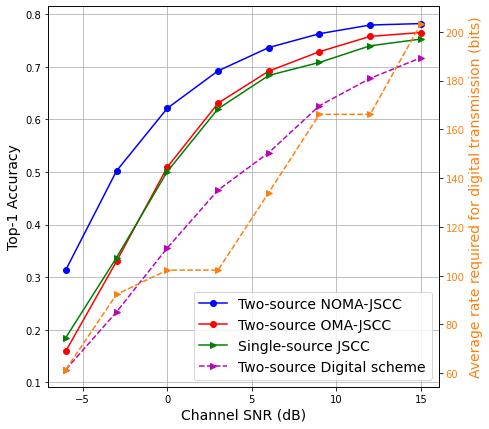}
         \caption{Slow fading channel with CSIR}
         \label{fig:top1_csi}
     \end{subfigure}%
        \caption{Top-1 retrieval accuracies of the proposed two-device schemes and the single-device scheme under different channel SNRs, with a total channel bandwidth of $q = 64$.}
        \label{fig:top1}
\end{figure*}

\section{ {Experimental Results}}
\subsection{Performance against channel SNR}
The proposed schemes for JSCC with OMA and NOMA are trained and tested on a pre-processed Market-1501 \cite{market1501} dataset over a wide range of channel SNRs from -6dB to 15dB, and compared with the separation-based scheme and the single-device JSCC scheme in \cite{jscc_ae}.

In Fig. \ref{fig:top1_awgn}, we plot the top-1 accuracy in an AWGN channel. In Fig. \ref{fig:top1_nocsi}, we plot the top-1 accuracy in a slow fading channel without CSI at the receiver. The digital scheme is not plotted in Fig. \ref{fig:top1_nocsi} because such a scheme is not possible to decode without CSI at the receiver, while JSCC allows communication even without the availability of CSI at the receiver. In Fig. \ref{fig:top1_csi}, we plot the top-1 accuracy in a slow fading channel with CSI available at the receiver. As expected, CSIR provides better accuracy than when CSI is absent at the receiver.

In Figs. \ref{fig:top1_awgn}, \ref{fig:top1_nocsi} and \ref{fig:top1_csi}, the proposed JSCC schemes outperform the separate digital scheme at almost all SNRs, except at high SNRs. However, note that we assume MAC capacity-achieving codes with equal rate allocation for each transmitter in this separate digital scheme, and therefore the reported performance of the digital scheme is not achievable in practice, particularly for the very low channel bandwidth of $q = 32$ per user considered here. The two-device JSCC schemes outperform the single-device JSCC scheme for a wide range of channel SNRs, especially higher SNRs, showing that incorporating two views of the same identity to make a collaborative decision at the edge server improves the retrieval performance. It is also observed in Fig. \ref{fig:top1_awgn}, \ref{fig:top1_nocsi} and \ref{fig:top1_csi} that JSCC with NOMA outperforms its orthogonal counterpart. In Fig. \ref{fig:top1_awgn}, it is shown that while the OMA JSCC scheme outperforms the single-device JSCC benchmark at most SNRs, they are surpassed by it at very low SNRs. This is because, in the low SNR regime, it is more beneficial to allocate all the channel resources to one transmitter to acquire the features from that one with sufficient quality for retrieval, rather than receiving very low quality features from two queries. However, the NOMA JSCC scheme brings the benefits of both schemes together, and outperforms both schemes at all SNRs. In Fig. \ref{fig:top1_csi}, the single-device JSCC as well as the proposed two-device JSCC schemes (both OMA and NOMA) outperform the separation-based scheme. These observations match our expectations. The suboptimality of separate source and channel coding used in the digital transmission scheme stems from two reasons. First of them is the usual suboptimality of separation in the finite blocklength regime. This was already observed in \cite{jscc_ae} for a point-to-point scenario. On the other hand, even in the infinite blocklength regime, separation becomes suboptimal when the two sources transmitted over the MAC are correlated. It is known that exploiting the correlation between the sources to generate correlated codewords at the encoders can strictly increase the end-to-end performance \cite{1056273}, \cite{Lapidoth:TIT:10}. To allow partial cooperation between the distributed transmitters, we must allow the transmitted signals to depend statistically on the source outputs, thus inducing correlation between the transmitted signals. Separation-based schemes operate in the opposite manner, where the dependence between the sources is destroyed by separate source and channel coding, thus making the transmitted signals independent. 
\begin{figure*}[!t]
    \begin{subfigure}[b]{0.33\textwidth}
         \centering
         \includegraphics[width=0.86\linewidth]{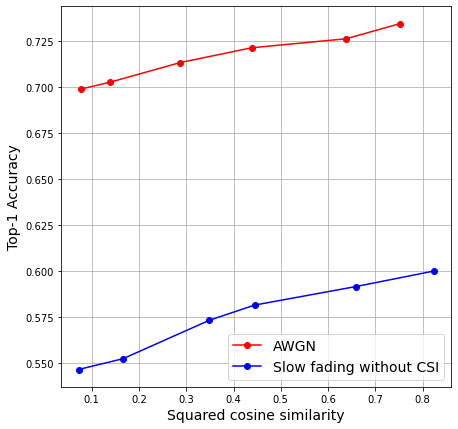}
         \caption{JSCC with NOMA - cosine similarity}
         \label{fig:cossim_acc}
     \end{subfigure}%
     \begin{subfigure}[b]{0.33\textwidth}
         \centering
         \includegraphics[width=0.83\textwidth]{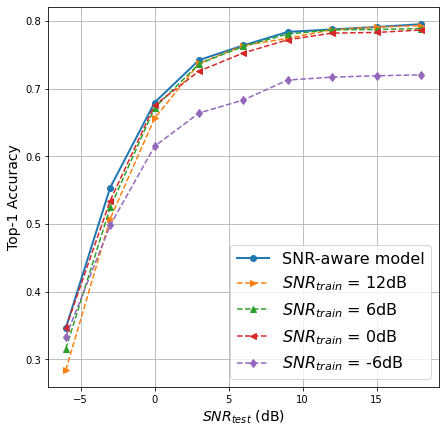}
         \caption{JSCC with OMA - AWGN channel}
         \label{fig:snr_ortho_awgn}
     \end{subfigure}%
     \begin{subfigure}[b]{0.33\textwidth}
         \centering
         \includegraphics[width=0.84\textwidth]{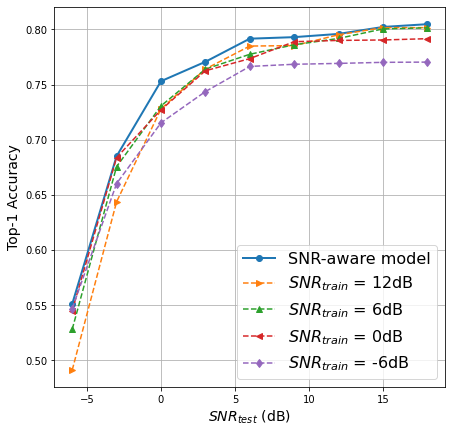}
         \caption{JSCC with NOMA - AWGN channel}
         \label{fig:snr_over_awgn}
     \end{subfigure}%
        \caption{Top-1 retrieval accuracies of: (a) the JSCC-NOMA scheme on AWGN and slow fading channels against different squared cosine similarity between $\mathbf{x}_1$ and $\mathbf{x}_2$, with channel $\text{SNR} = 0\text{dB}$, and channel bandwidth $q = 64$, and (b),(c) the SNR-aware scheme and the original schemes trained with various $\text{SNR}_{train}$ values against different $\text{SNR}_{test}$ values for OMA and NOMA schemes.}
        \label{fig:snr}
\end{figure*}

\begin{table}[!t]
    \centering
    \begin{tabular}{c|c}
    \textbf{Scheme} & \textbf{Squared cosine similarity}\\
    \hline
     OMA (AWGN) & 0.0151 \\
     OMA (slow fading) & 0.0165 \\
     NOMA (AWGN) & 0.7523  \\
     NOMA (slow fading) & 0.8234  \\ 
    \end{tabular}
    \caption{Squared cosine similarity between input symbols of the OMA and NOMA schemes.}
    \label{table:cos_sim_value}
\end{table}

We observe that the orthogonal JSCC architecture learns to transmit uncorrelated signals, as shown in Table \ref{table:cos_sim_value}, where the correlation between $\mathbf{x}_1$ and $\mathbf{x}_2$ is computed using squared cosine similarity, defined as $\cos^2(\mathbf{x_1},\mathbf{x_2}) \triangleq \frac{\langle \mathbf{x_1},\mathbf{x_2} \rangle^2}{\|\mathbf{x_1}\|^2\|\mathbf{x_2}\|^2}$. By sending independent symbols, the JSCC encoders capture non-overlapping information from the two views, thus avoiding redundancy, and maximising the use of communication resources. However, this mechanism is unable to make the distributed transmitters cooperate through the dependence of transmitted signals; hence, the lower accuracy achieved compared to the NOMA scheme. In contrast, JSCC with NOMA learns to transmit correlated signals. Higher correlation between the transmitted signals for the NOMA scheme results in higher performance. In fact, in Fig. \ref{fig:cossim_acc}, we plot the effect of the amount of correlation between the transmitted signals on the performance of the NOMA JSCC scheme, which we control by introducing a cosine similarity regularization term in the loss function as follows: $l = \frac{1}{3} (l_{ce_{aux1}} + l_{ce_{main}} + l_{ce_{aux2}}) + \lambda \cos^2{(\mathbf{x_1},\mathbf{x_2})}$.  Higher values of $\lambda$ force $\mathbf{x}_1$ and $\mathbf{x}_2$ to be less correlated. We observe in Fig. \ref{fig:cossim_acc} that the accuracy drops as the correlation between the transmitted signals decreases. Interestingly, when the cosine similarity in the NOMA scheme is reduced to approach $0$, its accuracy approaches that of the orthogonal JSCC scheme.

\subsection{SNR-aware JSCC}
The SNR-aware JSCC scheme, introduced in Section \ref{sec:jscc}, is trained over a range of $\text{SNR}_{train}$, and tested over a wide range of  $\text{SNR}_{test}$ values, from -6 to 18dB. In Fig. \ref{fig:snr_ortho_awgn} and \ref{fig:snr_over_awgn}, the performance of the SNR-aware schemes for the two JSCC schemes is compared with that of non-SNR-aware architectures trained over a single $\text{SNR}_{train}$ but tested on different $\text{SNR}_{test}$ values.

Note that the non-SNR-aware architectures exhibit graceful degradation when there is channel mismatch, that is, when the test channel conditions are worse than that of the training conditions. Thus, the JSCC scheme is able to avoid the cliff effect which conventional digital communication suffers from, where the performance of the digital schemes drops sharply when channel conditions are worse than those for which the encoder and decoder are designed. However, the SNR-aware architectures are observed to achieve strictly higher retrieval accuracies than the non-SNR-aware architectures (see Fig. \ref{fig:snr_ortho_awgn} and \ref{fig:snr_over_awgn}), providing a single DNN that performs the same or better on all SNRs than employing a distinct DNN optimised for each particular SNR value or range.

\section{Conclusion}
We proposed two JSCC schemes for deep-learning based distributed retrieval at the wireless edge, with OMA and NOMA, respectively. These schemes are shown to outperform conventional separation based alternative with capacity-achieving channel codes, and the JSCC scheme with a single source \cite{jscc_ae}. 
We observed that the NOMA JSCC scheme outperforms its OMA counterpart with TDMA. We also observed that the DNN architecture, when trained for NOMA, learns to transmit correlated signals to induce partial cooperation between the transmitters and to improve the final accuracy. The OMA JSCC scheme, in contrast, learns to transmit uncorrelated signals. 
With these observations in mind, in our future work, we will study how the correlation between the transmitted signals can be optimized to improve performance. 

\bibliographystyle{IEEEtran}
\bibliography{main}

\appendices
\section{Network architecture}
This section provides details of each component of our proposed method. Fig. \ref{fig:jscc} shows the whole pipeline of our method.

\subsection{JSCC network structure}
We use the JSCC method for feature transmission, where $\mathbf{v_1}$ and $\mathbf{v_2}$ are mapped into the channel symbols $\mathbf{x_1}$ and $\mathbf{x_2}$ by the JSCC encoders, and $\mathbf{\hat{v}_1}$ and $\mathbf{\hat{v}_2}$ are decoded by the JSCC decodes based on the received signals.

\begin{figure}[!h]
\begin{center}
   \includegraphics[width=1\linewidth]{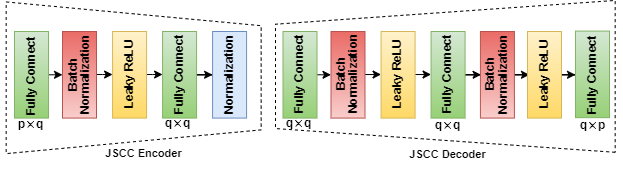}
   \end{center}
   \caption{DNN architecture for the JSCC transmission schemes.}
\label{jscc_general}
\end{figure}

Detailed structure of each JSCC encoder and decoder are shown in Fig \ref{jscc_general}, where $p$ is the length of the feature vectors, and $q$ is the available bandwidth of each transmitter.

\begin{figure*}[t]
\begin{center}
   \includegraphics[width=0.9\linewidth]{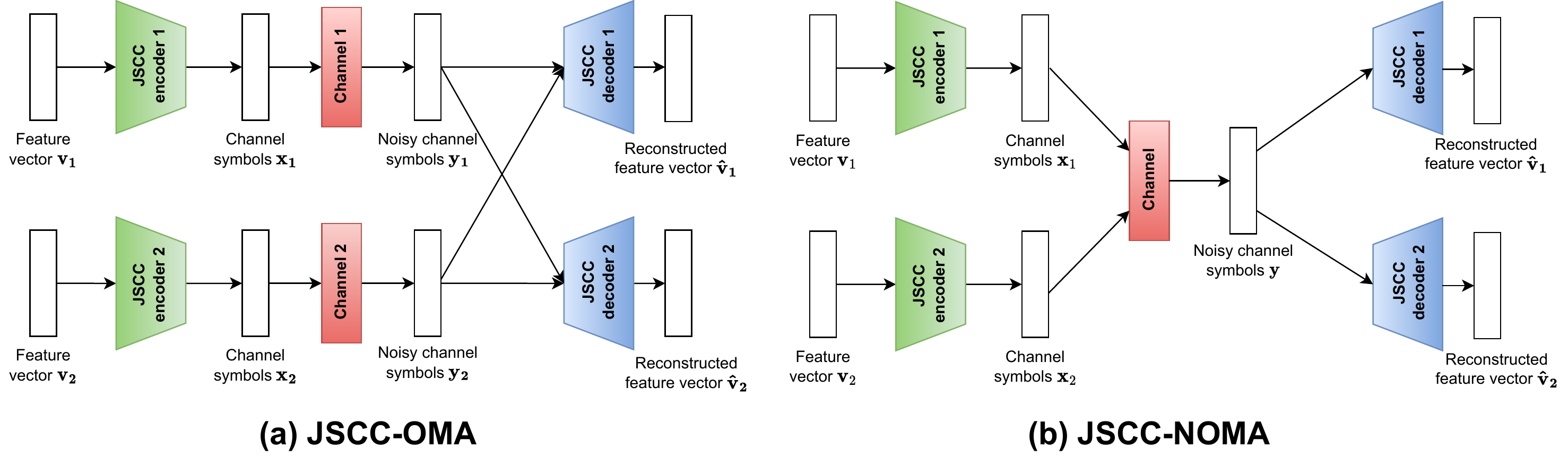}
   \end{center}
   \caption{JSCC model in JSCC-OMA and JSCC-NOMA}
\label{fig:jscc_model}
\end{figure*}

As we study both OMA and NOMA transmission schemes, we list the detailed comparisons of the JSCC model for them in Fig. \ref{fig:jscc_model}. For the OMA model, each transmitter is allocated half the available channel bandwidth, so we have $\mathbf{x_i}\in \mathbb{C}^\frac{q}{2}$, $\mathbf{y_i} = h_i\mathbf{x}_i+ \mathbf{z_i}$, $\mathbf{y_i}\in \mathbb{C}^\frac{q}{2}$, $\mathbf{y}\triangleq[\mathbf{y_1};\mathbf{y_2}]^T\in \mathbb{C}^q$, where $\mathbf{z_i} \in \mathbb{C}^\frac{q}{2}$ is the additive noise vector, and $i$ is the index of the transmitter from which the signal is received. For the NOMA model, each transmitter is allocated the total bandwidth, so we have $\mathbf{x_i}\in \mathbb{C}^q$, $\mathbf{y} = h_1\mathbf{x}_1 + h_2\mathbf{x}_2 + z$, $\mathbf{y}\in \mathbb{C}^q$, where $\mathbf{z} \in \mathbb{C}^q$ is the additive noise vector. 

\subsection{Details of Image Retrieval module}
The receiver uses an image retrieval module to identify the object's identities based on the received signals and the local database. 

Specifically, we use the same approach as that in the single-source model \cite{jscc_ae} during the training of the models. In the training stage, we add classification layers at the end of the network and use the cross-entropy loss Eqn. \eqref{eq:idloss} as part of the loss function for the model's training. In the testing stage, we discard the classification layers and perform the nearest neighbors search algorithm\cite{market1501,top-k} over the $\mathbf{\hat{v}_1}$ and $\mathbf{\hat{v}_2}$ to predict the identities of the objects, where the Euclidean distance between the received feature vectors of the query images, and the feature vectors of the images in the gallery database, is used to measure the similarity. More precisely, we compute the similarity between the query image pair and each gallery image, where the view-pooling layer can be treated as a concatenation operation before the similarity computation. Given the computed similarities, a rank list of gallery images is generated, which sorts the gallery images with identities in descending order of similarity with the query image pair. A pre-determined evaluation metric (such as top-1 accuracy\cite{market1501,top-k}) is then used on the rank list to evaluate the performance.

\textbf{Top-1 accuracy metric \cite{cmc}:} Given a query image, we compute a rank list of the gallery images sorted in descending order of their similarity to the query image pair. Then we account the top-1 accuracy\cite{top-k} $Acc_1$ for each query image as:
\begin{equation}
    Acc_{1} = \begin{cases}
    1 & \text{top-1 ranked image contains the query identity}\\
    0 & \text{otherwise}
    \end{cases}
\label{eq:top1}
\end{equation}

After the top-1 accuracy for every query image pair is computed, the final top-1 retrieval accuracy of the model is computed by averaging the $Acc_1$ for all the query image pairs. In other words, the top-1 accuracy of the model is the proportion of the correct identities at the top of the rank list for each query image pair.






\section{Details of the training strategy}
This section provides the details of our training strategy, including the loss functions and hyperparameters used. We employ a three-step training strategy, consisting of feature encoders pre-training ($\mathrm{T_1}$), JSCC autoencoders pre-training ($\mathrm{T_2}$), and end-to-end training ($\mathrm{T_3}$).

Similarly to the $\mathrm{T_1}$ phase in \cite{jscc_ae}, only two semantic feature encoders are pre-trained by three classifiers. During the $\mathrm{T_1}$ pre-training phase, each classifier predicts the identity over the features obtained from the semantic encoders. The average cross-entropy loss function is defined as follows: 
\begin{equation}
    l_{cls} = \frac{1}{3} (l_{ce_{aux1}} + l_{ce_{main}} + l_{ce_{aux2}}).
    \label{eq:idloss}
\end{equation}
where $l_{ce_{aux1}},l_{ce_{aux2}},$ and $l_{ce_{main}}$ are the cross-entropy loss terms computed over the identity prediction results from three classifiers, namely the auxiliary classifier 1, the auxiliary classifier 2, and the main classifier, respectively, and the one hot ground truth label $\mathbf{{r}}$. The classifiers consist of a fully-connected layer followed by a softmax operation, whose outputs are denoted by $\mathbf{\hat{r}_{1}}$, $\mathbf{\hat{r}_{2}}$, and $\mathbf{\hat{r}_{3}}$, respectively. The cross-entropy loss function\cite{personreid} is given as:

\begin{equation}
    l_{ce}(\mathbf{{y}},\mathbf{\hat{y}}) = -\sum_{i=1}^M {{y_{i}}}\log({\hat{y}_i}),
\end{equation}
where $M$ is the total number of classes, $\mathbf{{y}}$ is the ground truth of the identity (one-hot vector), and $\mathbf{\hat{y}}$ is the predicted result from a classifier, $y_{i}$ is the binary indicator from the one-hot vector, $\hat{y}_{i}$ is the predicted probability of the sample belonging to the class $i$. Then we have, $l_{ce_{aux1}}\triangleq {l}_{ce} (\mathbf{\hat{r}_{1}},\mathbf{r})$, $l_{ce_{aux2}}\triangleq {l}_{ce} (\mathbf{\hat{r}_{2}},\mathbf{r})$, and $l_{ce_{main}}\triangleq {l}_{ce} (\mathbf{\hat{r}_{3}},\mathbf{r})$.

For the training details, the feature encoders are trained for 30 epochs with a batch size of 16, and a learning rate of 0.01. Stochastic gradient descent (SGD) with a momentum of 0.9 is used as the optimizer, and an L2 regularizer weighted by $5 \cdot 10^{-4}$ is also applied. 


In $\mathrm{T_2}$, the pre-trained semantic feature encoders are frozen, and only the JSCC autoencoders are trained. The loss function used is the average mean-squared error (MSE) losses between the transmitted and reconstructed semantic features:
\begin{equation}
    l_{jscc} = \frac{1}{2} (l_{MSE_{1}} + l_{MSE_{2}}).
\label{eq:mse_avg}
\end{equation}
where $l_{MSE_{1}}\triangleq \text{MSE}(\mathbf{v_1},\mathbf{\hat{v}_1})$, $l_{MSE_{2}}\triangleq \text{MSE}(\mathbf{v_2},\mathbf{\hat{v}_2})$, and $\mathbf{v_i}$ and $\mathbf{\hat{v}_i}$ are the input and the output of the $i$-th JSCC encoder and JSCC decoder. The MSE loss is given as $ \text{MSE}(\mathbf{x},\mathbf{\hat{x}})\triangleq E[\|\mathbf{x}-\mathbf{\hat{x}}\|^2_2]$, where the expectation is taken over all pixels of the input pair $(\mathbf{x},\mathbf{\hat{x}})$. 

In our case, the network is trained for 200 epochs with an initial learning rate of 0.1, which is reduced to 0.01 after 150 epochs. The same optimizer and L2 regularization as applied as in $\mathrm{T_1}$.

In $\mathrm{T_3}$, the whole network is trained jointly with the loss function same as that in $\mathrm{T_1}$, as of Equation \ref{eq:idloss}. The whole model is trained for 30 epochs in total, with different learning rates for different components. For the feature encoder and the classifiers, the initial learning rate is 0.01, and it is reduced to 0.001 after 20 epochs. As for the JSCC autoencoder, the learning rate is 0.001 for the first 20 epochs and 0.0001 for the remaining 10 epochs. The same optimizer and L2 regularization as the previous steps are used.


\section{Detailed quantization and channel coding method of the digital scheme}
Many efficient channel coding methods, such as low-density parity-check and polar codes, can be used for the channel coding scheme. In our paper, we assume the channel capacity achieving channel codes as an upper bound for the digital scheme in the experiments of Section V, which is the same with \cite{jscc_ae}. Note that the channel capacity provides only an upper bound on the maximum reliable communication rate, and is not achievable in practice, particularly at the very short blocklengths considered here. 

For the detailed quantization methods, we utilize the well-known quantization noise to make the quantization process end-to-end and differentiable. Specifically, instead of rounding the latent representation to the nearest integer, in the training phase, we add the uniform noise to each element of the latent representation as follows:
\begin{equation}
    Q(\mathbf{v_i}) = \mathbf{v_i} + \mathcal{U}(-\frac{1}{2},\frac{1}{2})
\label{eq:quant}
\end{equation}
where $Q(\cdot)$ is the approximated quantization operation, $\mathbf{v_i}$ is the semantic feature vector from the source $\mathbf{s}_i$, and $\mathcal{U} (\cdot, \cdot)$ is the uniform noise vector. This formulation ensures a good approximation of quantization during training, whereas we perform rounding to the nearest integer during inference.

In the training phase, we also evaluate the average approximate entropy over the dataset in our loss function, which we define as a weighted sum of two objectives:

\begin{equation}
    l = \frac{1}{3} (l_{ce_{aux1}} + l_{ce_{main}} + l_{ce_{aux2}}) + \lambda \cdot (\log_2 p(\tilde{\mathbf{v}}_1) + \log_2p(\tilde{\mathbf{v}}_2)),
\label{eq:loss}
\end{equation}
where $l_{ce_{aux1}}, l_{ce_{main}}$, and $l_{ce_{aux2}}$ are the cross-entropy computation between the predicted class (identity) and the ground truth for the retrieval task, same to the \cite{fu2019horizontal,jscc_ae}. The second component of the loss function corresponds to the empirical Shannon entropy of the quantized vector, the same as the \cite{jscc_ae}, representing the average length of the output of the arithmetic encoder. Such loss formulation\cite{jscc_ae} allows for a smooth transition between the retrieval accuracy and the number of bits necessary to send the feature vector in a lossy fashion. 

\section{CSI-aware models}
For CSI-aware architecture, there are two different structures. One is a fading-aware model which adapts to channel fading gains to achieve channel equalization in the case of a slow fading channel with CSIR. The other is an SNR-aware model, which exploits the attention mechanism to cater to channel SNR mismatch between training and testing.

\subsection{AF module}
Inspired by \cite{wu2022channel,jscc_attention}, we introduce an attention feature (AF) module to help the model learn to adapt the given CSI.

The AF module scales the input feature dynamically with the available CSI in three steps: context extraction, factor prediction, and feature recalibration, with the output feature having the same dimension. In the context extraction step, the input vector $\mathbf{f}$ is concatenated with the corresponding channel state. The concatenated vector is fed as input to the factor prediction step, in which an attention mask $\mathbf{s}$ is predicted through a factor prediction neural network. This neural network is a simple network consisting of only two FC layers. Finally, at the feature recalibration step, the attention mask $\mathbf{s}$ is multiplied element-wise with the input vector $\mathbf{f}$ to give the scaled output vector $\mathbf{\hat{f}}$, which has the same dimension as the input vector. Through the above steps, each element of the input vector is scaled individually according to the corresponding channel gain. 

\subsection{Fading-aware model}
We design a fading-aware model for the CSIR scenario in the slow fading channel. Similar to \cite{wu2022channel}, our channel fading-aware model can adapt to the available CSI to achieve better results with the help of the attention module, compared with the original models with a traditional channel equalization operation. 

\begin{figure}[t]
\begin{center}
   \includegraphics[width=0.75\linewidth]{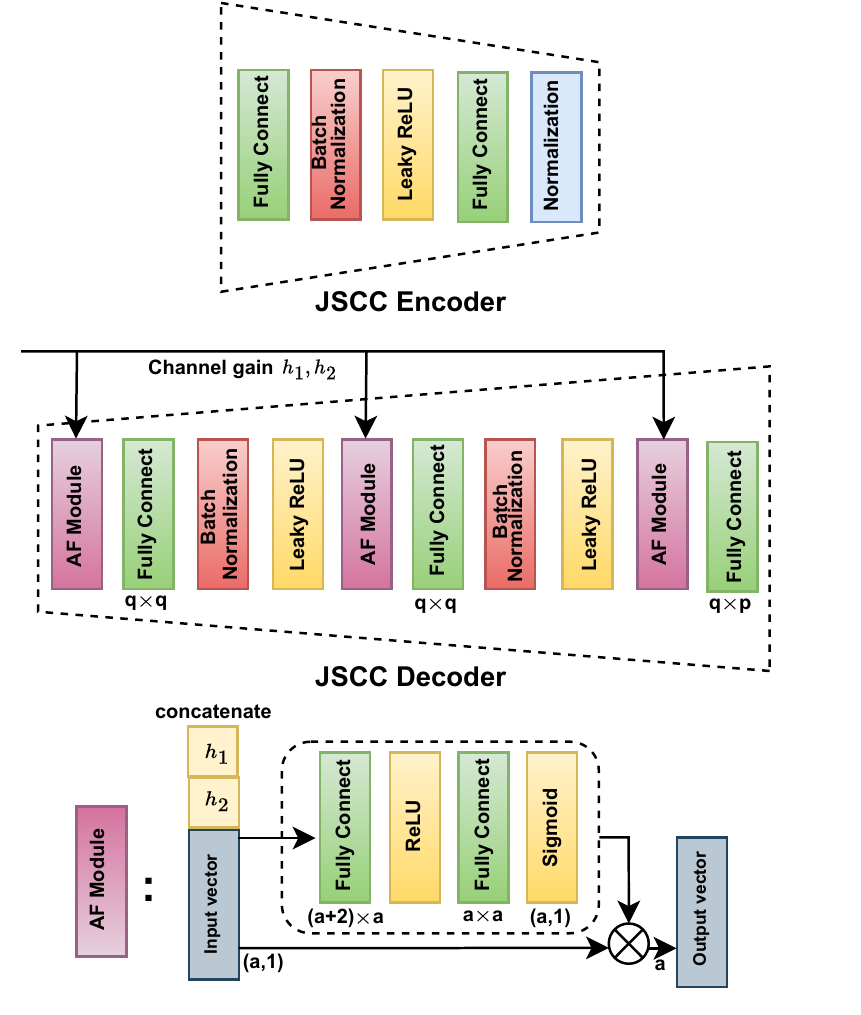}
   \end{center}
   \caption{Architecture of the channel fading-aware JSCC for slow fading channel with CSIR, where the channel gains are given into the attention mechanism, $a$ is the input dimension.}
\label{fig:jscc_csir}
\end{figure}

The architecture of our channel fading-aware model is shown in Fig. \ref{fig:jscc_csir}. Compared with a general JSCC (Fig. \ref{jscc_general}), the encoder is the same, and the decoder is different, where the available CSI is fed into the decoder via the AF modules inserted between each layer. The design of this model allows the network to learn to adapt the CSI in the decoder to achieve better performance via end-to-end training.

\subsection{SNR-aware model}
\begin{figure}[t]
\begin{center}
   \includegraphics[width=0.75\linewidth]{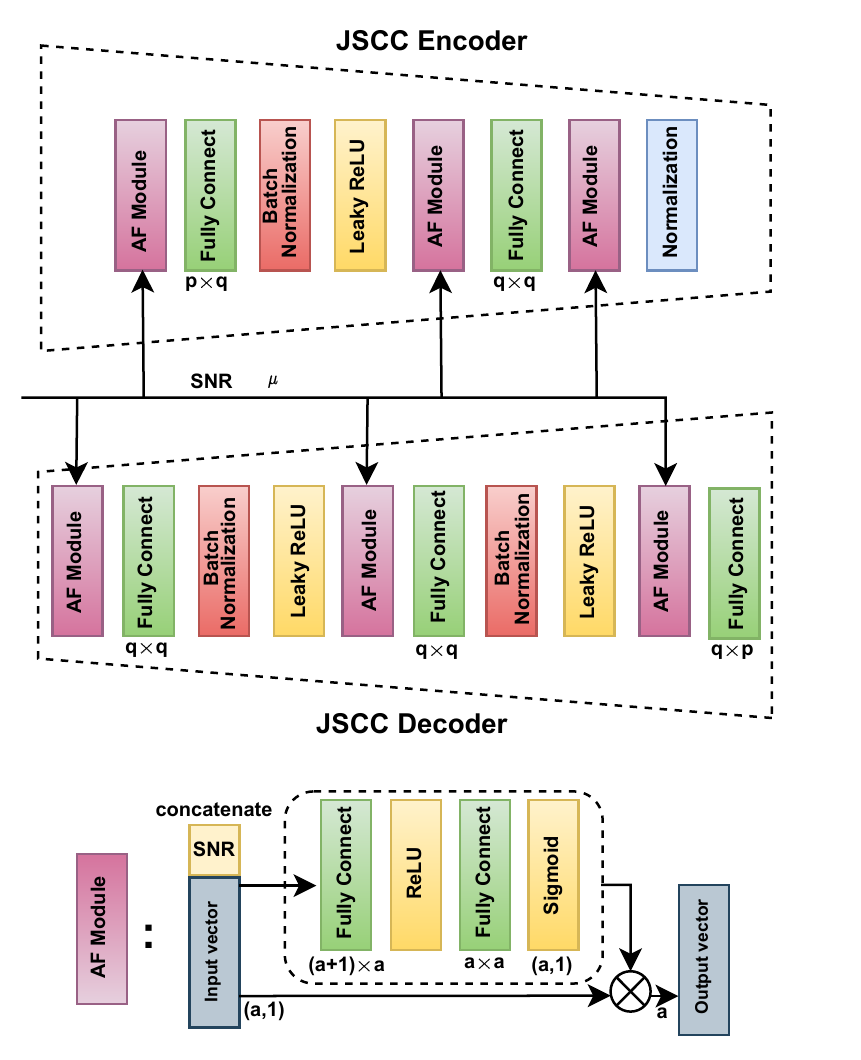}
   \end{center}
   \caption{Architecture of the SNR-aware JSCC, where the channel SNR is given into the attention mechanism, $a$ is the input dimension.}
\label{fig:jscc_ae}
\end{figure}

We design an SNR-aware model to enhance the original models in a way such that the models can cater to situations when there is a mismatch between the training and testing channel SNRs. Based on the previous success of the attention mechanism in JSCC for wireless image transmission \cite{jscc_attention,wu2022channel}, we exploit the AF module to scale the intermediate features dynamically based on the SNR. This way, the channel SNR will be fed into the network as one of the inputs, and the model will become adaptive to the various SNRs. The architecture of our SNR-aware model is shown in Fig. \ref{fig:jscc_ae}, where AF modules are inserted in between layers in the JSCC autoencoder, at which the channel SNR is fed into the modules, making it SNR adaptive.

\end{document}